\documentclass[aps,prb,twocolumn,preprintnumbers,amsmath,amssymb,superscriptaddress,floatfix]{revtex4-2}

\usepackage{graphicx}
\usepackage{dcolumn}
\usepackage{amsmath}
\usepackage{amssymb}
\usepackage{color}
\usepackage{multirow}
\usepackage{url}
\usepackage{booktabs}
\usepackage{tabularx} 

\usepackage[utf8]{inputenc}
\usepackage{siunitx} 
\usepackage{verbatim} 
\usepackage[normalem]{ulem}
\usepackage{placeins} 
\usepackage{float} 
\usepackage[hidelinks]{hyperref} 
\hypersetup{
colorlinks=true,    
linkcolor=blue,
citecolor=blue,
urlcolor=blue,
pdfborder={0 0 0},
bookmarksopen=true,
bookmarksopenlevel=2,
pdfpagemode=UseOutlines,
pdfstartview={Fit},
pdftitle={},
pdfauthor={Z. Salman},
pdfsubject={Advanced muon-spin spectroscopy},
pdfkeywords={},
pdfhighlight= /I
}
\newcommand{\tHMoTet}{2\textit{H}\protect\nobreakdash-{MoTe}$_2$}

\newcommand{\mSR}{$\mu$SR}
\newcommand{\bnmr}{$\beta$-NMR}

\newcommand{\Li}{$^8$Li}
\newcommand{\Lip}{$^8$Li$^+$}
\newcommand{\Tenmr}{${}^{125}$Te}

\newcommand{\TN}{$T_{\rm N}$}

\begin{document}
\title{Contrasting impurity-induced magnetism and dynamics in 2H-MoTe$_2$}

\author{Jonas~A.~Krieger}
\email{jonas.krieger@psi.ch}
\affiliation{PSI Center for Neutron and Muon Sciences, 5232 Villigen PSI, Switzerland}
\altaffiliation[Previous affiliation: ]{Max Planck Institut f\"ur Mikrostrukturphysik, Weinberg 2, 06120 Halle, Germany}
\author{Igor~P.~Rusinov}
\affiliation{Lebedev Physical Institute of the Russian Academy of Sciences, Moscow, Russia}
\author{Sourabh~Barua}
\altaffiliation[Current address: ]{
Department of Physics, Central University of Punjab, Bathinda 151401, India}
\affiliation{Department of Physics, University of Warwick,
Coventry CV4 7AL, UK}
\author{Aris~Chatzichristos}
\affiliation{ Department of Physics and Astronomy, University of British 
Columbia, Vancouver, BC V6T 1Z1, Canada}
\affiliation{Stewart Blusson Quantum Matter Institute, University of 
British 
Columbia, Vancouver, BC V6T 1Z4, Canada }
\author{Jared~Croese}
\affiliation{CERN, Esplanade des Particules 1, 1211 Geneva, Switzerland}
\author{Derek~Fujimoto}
\altaffiliation[Current address: ]{TRIUMF, Vancouver, British Columbia V6T 2A3, Canada}
\affiliation{ Department of Physics and Astronomy, University of British 
Columbia, Vancouver, BC V6T 1Z1, Canada}
\affiliation{Stewart Blusson Quantum Matter Institute, University of 
British 
Columbia, Vancouver, BC V6T 1Z4, Canada }
\author{Stefan~Holenstein}
 \affiliation{PSI Center for Neutron and Muon Sciences, 5232 Villigen PSI, Switzerland}
\affiliation{Physik-Institut der Universit\"at Z\"urich, 
Winterthurerstrasse}
\author{Victoria~L.~Karner}
\affiliation{Department of Chemistry, University of British Columbia, 
Vancouver, 
BC V6T 1Z1, Canada}
\affiliation{Stewart Blusson Quantum Matter Institute, University of 
British 
Columbia, Vancouver, BC V6T 1Z4, Canada }
\altaffiliation[Current address: ]{TRIUMF, Vancouver, British Columbia V6T 2A3, Canada}
\author{Ryan~M.~L.~McFadden}
\altaffiliation[Current address: ]{TRIUMF, Vancouver, British Columbia V6T 2A3, Canada}
\affiliation{Department of Chemistry, University of British Columbia, 
Vancouver, 
BC V6T 1Z1, Canada}
\affiliation{Stewart Blusson Quantum Matter Institute, University of 
British 
Columbia, Vancouver, BC V6T 1Z4, Canada }
\author{John~O.~Ticknor}
\affiliation{Department of Chemistry, University of British Columbia, 
Vancouver, 
BC V6T 1Z1, Canada}
\affiliation{Stewart Blusson Quantum Matter Institute, University of 
British 
Columbia, Vancouver, BC V6T 1Z4, Canada }
\author{W.~Andrew~MacFarlane}
\affiliation{Department of Chemistry, University of British Columbia, 
Vancouver, 
BC V6T 1Z1, Canada}
\affiliation{Stewart Blusson Quantum Matter Institute, University of 
British 
Columbia, Vancouver, BC V6T 1Z4, Canada }
\affiliation{TRIUMF, Vancouver, British Columbia V6T 2A3, Canada}
\author{Robert~F.~Kiefl}
\affiliation{ Department of Physics and Astronomy, University of British 
Columbia, Vancouver, BC V6T 1Z1, Canada}
\affiliation{Stewart Blusson Quantum Matter Institute, University of 
British 
Columbia, Vancouver, BC V6T 1Z4, Canada }
\affiliation{TRIUMF, Vancouver, British Columbia V6T 2A3, Canada}
\author{Geetha~Balakrishnan}
\affiliation{Department of Physics, University of Warwick,
Coventry CV4 7AL, UK}
\author{Evgueni~V.~Chulkov}
\affiliation{
Departamento de Pol\'{\i}meros y Materiales Avanzados: F\'{\i}sica, Qu\'{\i}mica y Tecnolog\'{\i}a,
Facultad de Ciencias Qu\'{\i}micas, Aptdo. 1072, 20018, San Sebasti\'an, Espa\~{n}a}
\affiliation{Tomsk State University, pr. Lenina 36, 634050 Tomsk, Russia}
\affiliation{St. Petersburg State University, Universitetskaya nab. 7/9, 
199034 St. Petersburg, Russia}
\author{Stuart~S.~P.~Parkin}
\affiliation{Max Planck Institut f\"ur Mikrostrukturphysik, Weinberg 2, 06120 Halle, Germany}
\author{Zaher~Salman}
\email{zaher.salman@psi.ch}
\affiliation{PSI Center for Neutron and Muon Sciences, 5232 Villigen PSI, Switzerland}

\begin{abstract}
We investigate the behavior of interstitial \Lip\ implanted near the surface of \tHMoTet\ using $\beta$-detected NMR. We find that, unlike the muon, \Lip\ does not show any signature of induced magnetism. This result is consistent with density functional theory, which identifies the Li stopping site at the 2a Wyckoff position in the van der Waals gap and confirms the absence of detectable Li-induced electronic spin polarization. Both the spin-lattice relaxation and the resonance lines show evidence of strong spin dynamics above $\sim\SI{200}{K}$, reminiscent of local stochastic \Lip\ motion within a cage. The resonance line shape consists of quadrupolar satellites on top of a broad central peak. To better understand the interaction of \Lip\ with the host material, we employ a frequency-comb measurement, by simultaneously exciting four frequencies corresponding to the first-order quadrupolar satellite transitions, $\nu_0 \pm 3\nu_{\mathrm{comb}}$ and $\nu_0 \pm\nu_{\mathrm{comb}}$ around the Larmor frequency $\nu_0$ as a function of $\nu_{\mathrm{comb}}$. This offers an enhanced sensitivity to the quadrupolar split portion of the line. Using this method, we find a small decrease of the quadrupolar frequency with increasing temperature, showing the typical behavior associated with thermally excited phonons and the absence of any magnetic response which was observed with other defects in \tHMoTet.
\end{abstract}

\maketitle

\section{Introduction}\label{sec:Intro} 
Transition metal dichalcogenides (TMDs) are a family of layered van der Waals (vdW) materials consisting of stacks of MX$_2$ trilayers, where the transition metal M is surrounded by six chalcogenide atoms X, usually either with a trigonal prismatic (\textit{H}) or distorted octahedral (\textit{T}) coordination. Despite this similarity in atomic structure, these materials exhibit a plethora of different physical phenomena, ranging from metallic, semiconducting, Weyl-semimetallic and superconducting, to ferro- and antiferromagnetic and charge density wave behavior~\cite{Manzeli2017}. 
For this reason they offer a considerable potential for vdW heterostructure 
engineering~\cite{Geim2013,Rajamathi2017}. 
Semiconducting TMDs are of particular interest for integration into electronic and optical applications,  for example, polymorphic MoTe$_2$, which in the hexagonal 2\textit{H}-phase has an indirect $\sim\SI{1}{eV}$ bandgap~\cite{Grant1975,Conan1979}. 
Interestingly, in the monolayer the bandgap becomes direct, which has been exploited to build transistors and
photo-detector devices~\cite{Lezama2014,Xu2015,Ding2018}.
Furthermore, \tHMoTet\ was recently predicted to be an obstructed atomic insulator, having Wannier charge centers located away from atomic positions~\cite{Li_obstructed_2022}. As a result it is expected to  host obstructed surface states on non-basal surfaces and line defects, associated with a high catalytic activity useful for hydrogen evolution reaction~\cite{Li_obstructed_2022}.

Semiconducting TMDs are further prone to host magnetic impurities, e.g., at defects or interstitials and in particular in the monolayer limit (though not exclusively). In \tHMoTet\ magnetic impurities can occur at anti-site defects~\cite{Guguchia2018},  hydrogen and transition metal doping of the monolayer~\cite{Ma2011,GonzalezHerrero2016} and on hydrogen-like muon impurities in the vdW gap~\cite{Krieger2023PRM}. Defects have also been known to strongly affect the  optical, electronic and catalytic properties~\cite{Lin2016,Li2016,McDonnell2022}.
Here we present beta detected nuclear magnetic resonance (\bnmr) data on spin-polarized \Lip\ implanted into a \tHMoTet\ single crystal. In these experiments \Lip\ acts as a charged impurity and is likely to be located in the vdW gap. Therefore, one might expect it to show a similar magnetic response to that of an implanted $\mu^+$. However, while we confirm the \Li\ site to be in the vdW gap, we detect no indications of magnetism, neither intrinsic nor induced. On the other hand, we do find a strong increase in the spin-lattice relaxation above $\sim\SI{200}{K}$ which points to the onset of motional dynamics of the \Lip\ above this temperature. We further demonstrate that a frequency-comb detection scheme can be used to accurately determine the quadrupolar splitting in the presence of an additional overlapping broad peak contributing to the resonant line. These measurements demonstrate the diverse response of \tHMoTet\ and other semiconducting TMDs to impurities and defects, which depend strongly on their specific interaction with the host material.

\section{Experiment}\label{sec:Exp} 
\begin{figure*}
\begin{center}
  \includegraphics[width=.95\linewidth]{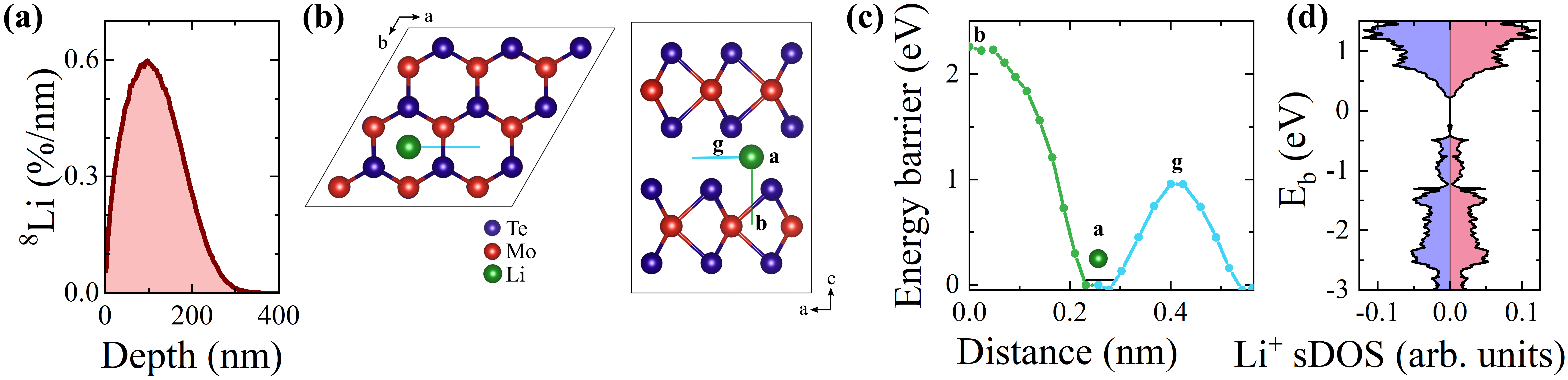}
\end{center}
\caption{
  \textbf{(a)} Simulated stopping profile of  $\SI{22.5}{keV}$ \Lip\ ions in \tHMoTet.
  \textbf{(b)} Stopping site and diffusion paths
\textbf{(c)} Diffusion energy barriers. The high symmetry sites along the \Lip\ diffusion path have been labeled by their Wyckoff letters of space group~194. 
(d) \Lip\ spin-DOS showing absence of induced magnetization.}
\label{fig:calc}
\end{figure*}

High-quality single crystals of \tHMoTet\ were grown by chemical vapor
transport, using TeCl$_4$ as a transport agent~[\onlinecite{Krieger2023PRM}]. The crystal studied here was taken from the same batch as those studied in Ref.~[\onlinecite{Krieger2023PRM}].
It was glued onto a sapphire substrate using silver paint and then mounted on a cold-finger cryostat. The surface of the crystal was cleaved initially, but kept in air for an extended period of time afterwards.
The \bnmr\ measurements were performed on the $\beta$-NQR~(Ref.~[\onlinecite{Zaher2006}]) and $\beta$-NMR~(Ref.~[\onlinecite{Morris2004PRL}])
spectrometers at the isotope separator and accelerator (ISAC) facility in TRIUMF,
Vancouver, Canada~\cite{Morris2014,Levy2014,Macfarlane2015,Macfarlane2022}.  Spin-polarized  \Lip\
 (spin $I=2$ and quadrupole moment $Q\sim\SI{32}{mb}$ with $\sim\SI{70}{\percent}$~polarization~\cite{Hatakeyama2002}) 
was implanted with an energy of $\SI{22.5}{keV}$ along the $c$-axis of the crystal. The corresponding ion-implantation stopping profile was simulated with the \texttt{Trim.SP}~\cite{Eckstein1991} software (using mass density of \SI{7.78}{g/cm\cubed}) and is shown in Fig.~\ref{fig:calc}(a). 
We find that the average implantation depth of \Lip\ is $\sim\SI{110}{nm}$. Note that the size of the crystal ($\sim\SI{25}{mm^2}$) was large compared to the beam spot diameter ($\sim\SI{2}{mm}$), thus we expect no significant background signal from ions missing the sample. The spin-lattice relaxation rate (SLR) was measured in an in-plane magnetic field,  $B\perp c$, and the line shape and frequency-comb measurements were performed in the presence of an out-of-plane field, $B\parallel c$.  In both cases, the fields were applied parallel to the initial \Lip\ nuclear spin direction. \Li\ is unstable, with a lifetime of $\tau_{^8\mathrm{Li}}=\SI{1.2}{s}$ and has an anisotropic $\beta$-decay, with the preferential emission direction of the decay electron being opposite to the direction of the \Li\ nuclear spin~\cite{Morris2004PRL,Zaher2006}. This allows the measurement of the average spin direction of a \Li\ ensemble via the detection of the anisotropic decay electrons. 

The \Li\ SLR was measured by implanting a \SI{4}{second} pulse of \Lip\ beam while monitoring the asymmetry in the emitted electrons during and after the beam pulse.  The measured $\beta$-asymmetry is proportional to the time evolution of the average nuclear spin polarization and therefore reflects the SLR rate of \Li.  

The \Li\ resonance lines were measured by implanting \Li\ continuously and monitoring the decay asymmetry while sweeping a continuous wave (CW) radio frequency (RF) magnetic field through the resonance~\cite{Morris2004PRL}. A quadrupole split resonance due to an electric field gradient (EFG) at the \Li\ stopping site appears in the spectra as a set of four small satellite lines in the vicinity of the Larmor frequency, $\nu_{\mathrm{L}}$.
This splitting was measured by using a frequency-comb instead of a single frequency~\cite{Minamisono1993,Adelman2022}.
This dramatically enhances the signal corresponding to the quadrupole satellites because the CW comb excites all four quadrupolar transitions simultaneously: $\nu_0 \pm\nu_{\mathrm{comb}}$ and $\nu_0\pm3\nu_{\mathrm{comb}}$, where $\nu_0$ is the predetermined center of the line and $\nu_{\mathrm{comb}}$ is the swept frequency parameter.  A depolarization of the \Li\ spin ensemble due to the first order quadrupolar transitions will therefore occur at the frequencies $\nu_{\mathrm{comb}}=\nu_q/3$, $\nu_{\mathrm{comb}}=\nu_q$, and $\nu_{\mathrm{comb}}=3\nu_q$, where the quadrupolar frequency $h\nu_q = eQV_{zz}/8$ arises from the  nuclear electric quadrupole moment $eQ$ and the EFG at the \Li\ site, of which $V_{zz}$ is the largest principal component. Note that on resonance (i.e., at $\nu_{\mathrm{comb}}=\nu_q$ and $\nu_{0}=\nu_{\mathrm{L}}$), the depolarization of the \Li\ ensemble will be further enhanced, since multiple quadrupolar transitions can be excited simultaneously~\cite{Stockmann1974,Ackermann1972}. The center frequency $\nu_0$ is chosen based on a preceding calibration measurement of the Larmor frequency. Therefore, it will only be accurate up to an error $\delta\nu_0$. As a consequence, the quadrupolar transitions will split and appear to be at $\nu_{\mathrm{comb}}=\nu_q/3\pm\delta\nu_0/3$, $\nu_{\mathrm{comb}}=\nu_q\pm\delta\nu_0/3,\pm\delta\nu_0$, and $\nu_{\mathrm{comb}}=3\nu_q\pm\delta\nu_0$. When the quadrupolar interaction is much smaller than the Zeeman energy in the applied field, this splitting is symmetric around the nominal transition frequency. Hence, as long as $\delta\nu_0$ is smaller than the width of the quadrupolar satellites, this splitting will not be visible and the effect of  $\delta\nu_0$ can be neglected without degrading the accuracy of the determined $\nu_q$. In these experiments, except for the measurement at RT, we are in this limit. It is important to point out here that $\delta\nu_0$ will affect in a nontrivial way the width and height of the peaks measured with the frequency-comb, making this approach less suited for an accurate determination of the \Li's initial state population. All measurements (SLR, resonances and comb) have been performed while switching the initial polarization direction and, where applicable, the RF sweep directions. The resulting

\section{DFT Calculations}
The \Li\ site and diffusion barriers were calculated with Density Functional Theory (DFT), consistent with the calculations used to determine the muon site in Ref.~[\onlinecite{Krieger2023PRM}].
Lithium was introduced in a relaxed 3$\times$3$\times$1 supercell of \tHMoTet, wherein we performed nudged elastic band (NEB)~\cite{Henkelman2000} calculations on the basis of the Linear Combination of Pseudo Atomic Orbitals (LCPAO) approach, as implemented in the OPENMX code~\cite{PAO}. A Perdew-Burke-Ernzerhof (PBE) exchange-correlation functional~\cite{PBE} has been applied. 
The Projected Augmented Wave (PAW) method from the VASP code~\cite{Kresse1996,Kresse1999} was used, taking into account both spin-orbit coupling  and scalar-relativistic corrections. 
The lithium ground state energies were estimated by approximating the 
NEB diffusion energy barriers with an anisotropic harmonic oscillator potential. The crystal 
structures were visualized using \texttt{VESTA}~\cite{VESTA}.

We find the stable \Lip\ site to be at the 2a Wyckoff position (Fig.~\ref{fig:calc}(b)). This site has a D$_{\rm{3d}}$ point group symmetry, so a nonzero EFG is expected which would result in a splitting of  the \Li\ quadrupolar transitions.
Potential charged states of the \Li\ are accounted for by assuming a neutral/charged 
supercell~\cite{Moeller2013,Bernardini2013,Krieger2023PRM}. We find the same 2a stopping site and the absence of induced spin polarization at the \Li\ site (Fig.~\ref{fig:calc}{(d)}) for both charged states.
The resulting calculated diffusion energy barriers along the paths indicated in  Fig.~\ref{fig:calc}{(b)} are shown in Fig.~\ref{fig:calc}(c).
We note that in MoS$_2$ the diffusion of Li between 2a sites was found to occur via a tetrahedral site above a Mo-atom~\cite{Shuai2016,Enyashin2012} and a similar stable Li site has also been considered in 2H-MoTe$_2$~\cite{Panda2020}. 
However, in our calculations, a diffusion path via the 6g site in between two Te atoms is found to be preferable. 

\section{Results}\label{sec:Res} 
\begin{figure*}
\begin{center}
  \includegraphics[width=.9\linewidth]{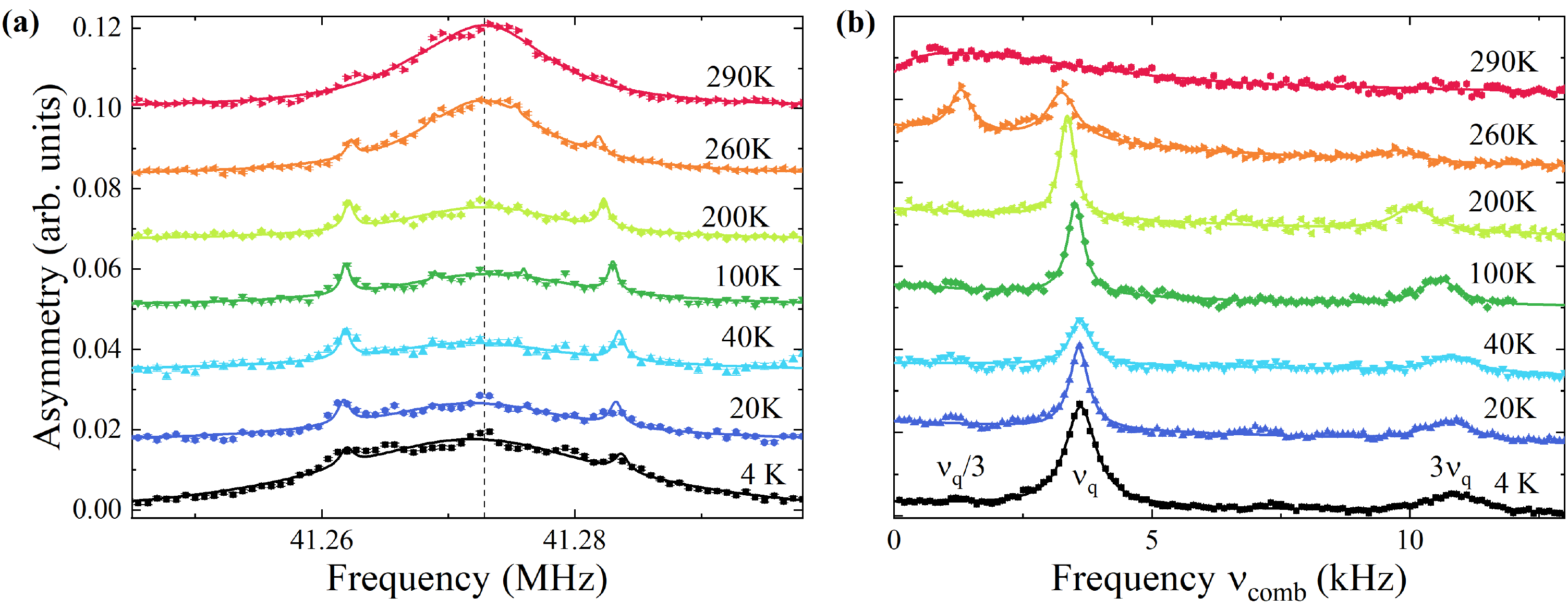}
\end{center}
\caption{
  \textbf{(a)} Typical \Li\ \bnmr\ line shapes for \tHMoTet\ in a field of \SI{6.55}{T}~$\parallel c$-axis. 
The spectra have been vertically offset for clarity. The solid lines show fits to the data and the vertical dashed line the position of the MgO reference resonance.
\textbf{(b)} \Li\ \bnmr\ spectra \tHMoTet\ at different temperatures in a field of \SI{6.55}{T} as a function of $\nu_{\rm{comb}}$ when using a frequency-comb to excite multiple quadrupolar transitions simultaneously, as described in the text with $\nu_0$ determined by fitting the line shapes in \textbf{(a)}. }
\label{fig:bnmrLines}
\end{figure*}

Typical \Li\ \bnmr\ line shapes measured in a \SI{6,55}{T} out-of-plane field at different temperatures are shown in 
Fig.~\ref{fig:bnmrLines}{(a)}.
They consist of a broad central resonance at all temperatures with small quadrupolar satellites appearing below \SI{260}{K}. Above this temperature, the quadrupolar splitting cannot be resolved.
In order to accurately determine the  satellite frequencies we performed frequency-comb measurements, representative spectra of which are shown in Fig.~\ref{fig:bnmrLines}(b). Except near room temperature, the peaks at $\nu=\nu_q$ and $\nu=3\nu_q$ are always clearly visible, whereas the peak at $\nu=\nu_q/3$ is only visible in  the vicinity of \SI{260}{K}. We have extracted the value of $\nu_q$ by fitting the frequency-comb spectra to a sum of two or three Lorentzian peaks, depending on whether the signal at $\nu=\nu_q/3$ was clearly resolved or not. The baseline from the central peak was phenomenologically parameterized by a second order polynomial, which gives a good agreement with the observed spectra, as shown by the solid lines in Fig.~\ref{fig:bnmrLines}(b). We then analyzed the resonance line-shapes [Fig.~\ref{fig:bnmrLines}(a)] by fitting them to a sum of five Lorentzian peaks, a central one and four corresponding to the quadrupolar splitting. The position and width of the latter was assumed to be given by the values extracted from the fits to the comb spectra. At the two highest temperature points, where no clear quadrupolar splitting is visible in the lineshape or frequency-comb, a single Lorentzian was assumed.

\begin{figure}
\begin{center}
  \includegraphics[width=.95\linewidth]{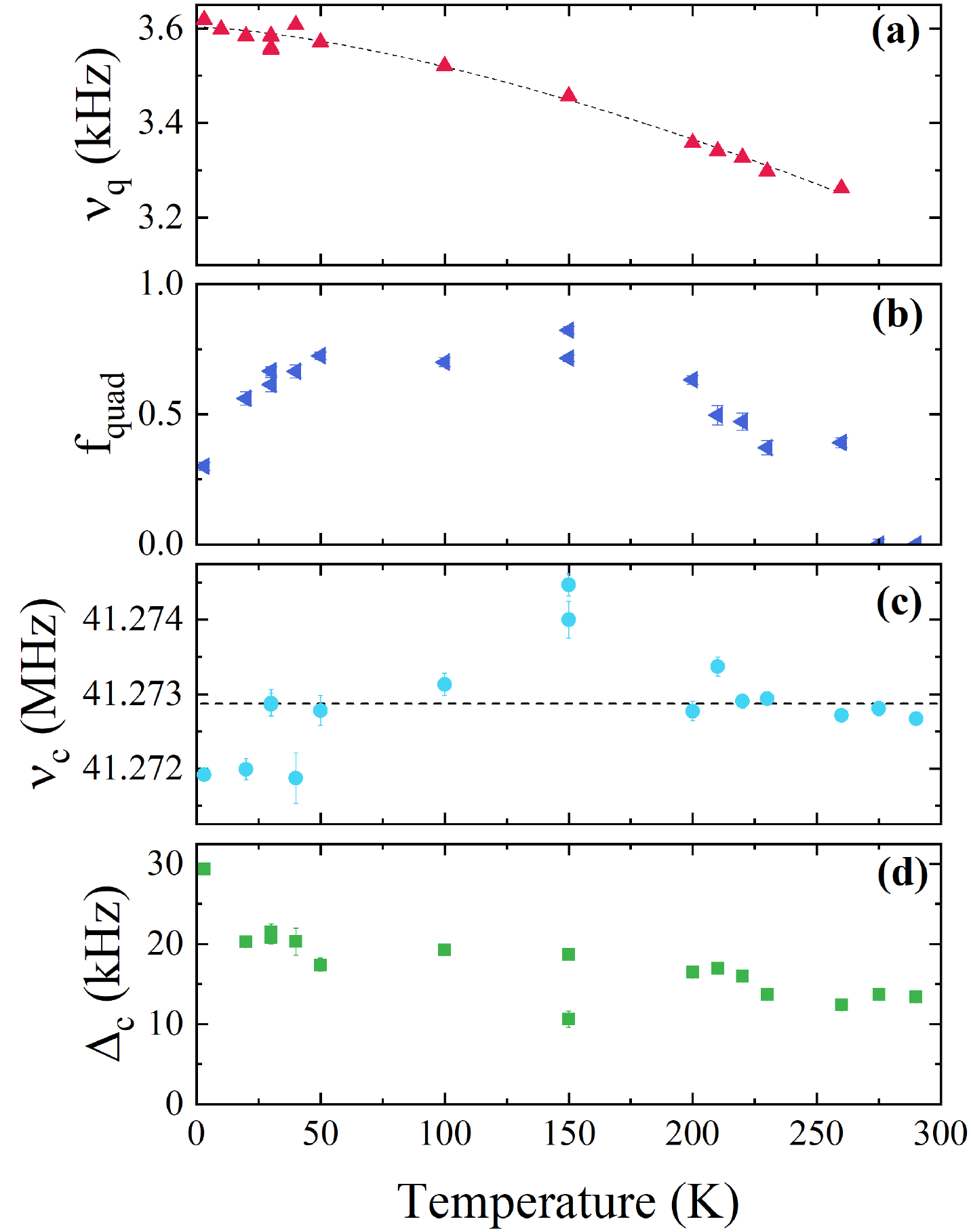}
\end{center}
\caption{Temperature dependence of the 
  \textbf{(a)} quadrupolar frequency,
  \textbf{(b)} relative amplitude of the quadrupolar peaks, 
  \textbf{(c)} resonance frequency,  and
  \textbf{(d)} width of the central line. The dashed line in \textbf{(a)} shows a $T^{3/2}$ power law behavior, and the dashed line in \textbf{(c)} shows the reference frequency of MgO. }
\label{fig:lineparams}
\end{figure}

The resulting temperature dependence of the resonance line parameters is shown in Fig.~\ref{fig:lineparams}.
The temperature dependence of the quadrupolar splitting in Fig.~\ref{fig:lineparams}{(a)} exhibits a power law behavior $\nu_q=\nu_{q0}(1-cT^{3/2})$, where $\nu_{q0}=\SI{3.603(4)}{kHz}$ and $c=\SI{2.33(7)e-5}{kHz/K^{3/2}}$ were determined via weighted least-squares minimization.
The resulting curve is shown as a dashed line in Fig.~\ref{fig:lineparams}{(a)}. Phenomenologically, this is the typical temperature dependence of the quadrupolar frequency in the presence of thermally activated phonons~\cite{Christiansen1976,Nikolaev2020}.
The quadrupolar contribution to the spectral weight was estimated by the sum of amplitudes of the four quadrupolar peaks relative to the sum of all five peaks and is shown in  Fig.~\ref{fig:lineparams}{(b)}. Note that the central peak frequency $\nu_c$  shows only little temperature dependence~[Fig.~\ref{fig:lineparams}{(c)}], with its full width at half maximum, $\Delta_c$, increasing with decreasing temperature~[Fig.~\ref{fig:lineparams}{(d)}].

\begin{figure*}
\begin{center}
  \includegraphics[width=.9\linewidth]{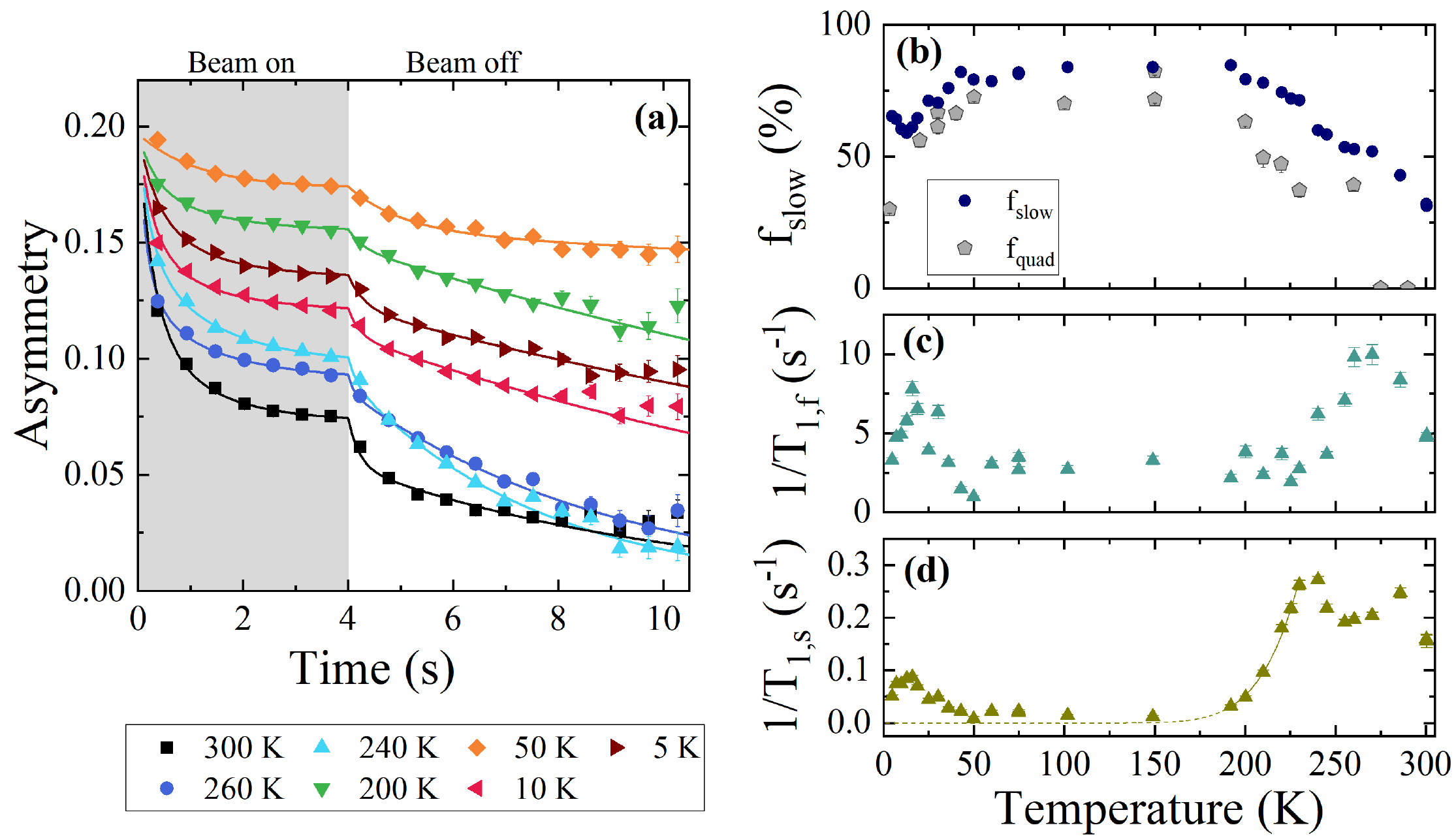}
\end{center}
\caption{ 
\textbf{(a)} Typical \Li\ SLR data measured in \tHMoTet. The \Li\ beam is implanted for \SI{4}{s} and then turned off (indicated by a gray shaded region). The solid lines show fits to the data. The temperature dependence of the slowly relaxing fraction \textbf{(b)}, the fast SLR ($1/T_{1{\text{,f}}}$) \textbf{(c)} and slow SLR ($1/T_{1{\text{,s}}}$) \textbf{(d)}.  The solid line shows a fit with an activation energy of $\sim\SI{0.2}{eV}$. For comparison, we also show in \textbf{(b)} also the relative amplitude of the quadrupolar peaks, $f_{\mathrm{quad}}$, from Fig.~\ref{fig:lineparams}{(b)}.
    }
\label{fig:bNMR}
\end{figure*}

Typical \Li\ SLR curves as a function of time in the presence of a \SI{10}{mT} in-plane field are shown in Fig.~\ref{fig:bNMR}{(a)}.
The relaxation of the \Lip\ ensemble occurs on a time scale of seconds, with distinct fast and slow relaxing contributions. From the raw data, the relaxation seems to be slowest for the curve at \SI{50}{K}. The time dependence of the SLR curves in Fig.~\ref{fig:bNMR}{(a)} can be described by a 
sum of two exponential depolarization functions, representing a fast ($1/T_{1{\text{,f}}}$) and a slow  ($1/T_{1{\text{,s}}}$) relaxing component:
\begin{equation}\label{eq:LiAsy}
 A(t)=A_0 \left(f_{\mathrm{slow}}e^{-t/T_{1{\text{,s}}}} +(1-f_{\mathrm{slow}})e^{-t/T_{1{\text{,f}}} }
 \right),
\end{equation}
where $f_{\mathrm{slow}}$ is the contribution of the slow relaxing fraction. The initial asymmetry $A_0$ 
was assumed temperature independent. Eq.~\ref{eq:LiAsy} was then convoluted with the duration of the beam pulse~\cite{Zaher2006}. Such a bi-exponential behavior is the simplest model that can adequately describe the data\footnote{For example, a stretched exponential $A(t)=A_0 e^{-\left(t/T_1\right)^\beta}$ fails, because the stretching coefficient 
$\beta$ converges to a non-physical values of $\beta\sim0.2$.}, while avoiding 
over-parametrization of the fits. 

The relaxation rates obtained from the fits are shown in Fig.~\ref{fig:bNMR}{(c,d)} and exhibit two peaks: One around \SI{240}{K} and the second at $\sim\SI{15}{K}$, the latter coinciding with a local minimum in $f_{\mathrm{slow}}$ [Fig.~\ref{fig:bNMR}(b)].
Note that any fluctuations coupling to the \Li\ spin and passing through the Larmor frequency ($\sim$\SI{63}{kHz} at \SI{10}{mT}) as a function of temperature could produce such peaks. An Arrhenius law was used to fit the low-temperature flank of the \SI{240}{K} SLR peak, giving an activation energy $E_A\sim\SI{0.2}{eV}$, as shown in Fig.~\ref{fig:bNMR}{(c)}.
This increase in SLR with the observed energy scale could be attributed to the onset of \Li\ diffusion, as previously reported in other vdW materials~\cite{McFadden2019}. Such a mechanism would result in dynamic changes of the EFG experienced by \Li, coupling to its quadrupolar moments and thereby enhancing its SLR. However, in this case, it is inconsistent with the absence of motional narrowing in the corresponding line-shapes, Fig.~\ref{fig:bnmrLines}{(a)}. In addition, the extracted activation energy of \SI{0.2}{eV} is much smaller than the estimated hopping barrier of $\approx\SI{1}{eV}$ predicted from DFT, Fig.~\ref{fig:calc}{(c)}. A possible explanation for this temperature dependence is discussed below.

\section{Discussion}\label{sec:Disc} 
Although the two components in the line shape could suggest that there are two inequivalent Li-states present in the sample, there are several indications that this is not due to two different crystallographic \Li\ stopping sites. On the theoretical side, the calculations only identify one candidate site, and there are no Wyckoff positions with a cubic point group symmetry in space group 194 that could explain the absence of an EFG on the broad central peak. Experimentally, the broad central line has a similar width to the total quadrupole splitting $3\nu_q$, which points to a scenario where an additional broadening affects the quadrupolar splitting.
But, more importantly, the quadrupolar fraction in Fig.~\ref{fig:lineparams}{(b)} seems to be strongly correlated to the slow fraction of the SLR [Fig.~\ref{fig:bNMR}(b)], which indicates that the absence of a detectable quadrupolar splitting is related to the presence of fast dynamics. The exact microscopic mechanism of the two \Li\ states is currently unknown. Possibilities include diffusional dynamics in a local cage, or  multiple \Li\ valence (charged) states with different sensitivity to fluctuations. For example, in addition to \Lip, a neutral \Li\ could be present \cite{MacFarlane2025PRB}. Long-range \Li\ diffusion is highly unlikely since we observe no motional narrowing at high temperature while the activation energy is much lower than that calculated for hopping between two neighbouring \Li\ sites.
If the bound electron in the neutral state corresponds to a very shallow donor state, then the associated large electronic orbital might explain the absence of a pronounced hyperfine shift.
Note that both a small hyperfine coupling and a relatively small quadrupolar splitting are typical for \Lip\ in the vdW gap of non-magnetic layered compounds~\cite{Wang2006,McFadden2019,Mcfadden2020,Ticknor2025UBC}.

In light of the magnetic state observed by \mSR~(Refs.~[\onlinecite{Guguchia2018,Krieger2023PRM}]), it is important to consider if there are similar magnetic signatures detectable with \bnmr.  However, the behavior of the \Li\ SLR at low temperature  is inconsistent with a strong local static magnetic field being present at the \Li\ position. Such a local field would depolarize the \Li\ spin completely on a timescale of milliseconds. 
The size of the local magnetic field in \mSR\ at low temperature is~$\sim\SI{200}{mT}$~\cite{Guguchia2018} and is associated with a magnetic muonium state at the 2a Wyckoff position~\cite{Krieger2023PRM}, which is the same as the \Lip\ stopping site. Given the smaller gyromagnetic ratio of \Li\ (${\gamma_{{}^8\mathrm{Li}}}/2\pi=\SI{6.3016}{MHz/T}$, with ${\gamma_{{}^8\mathrm{Li}}}/{\gamma_\mu}\sim\SI{4.7}{\percent}$), a similar local field would still cause a precession with $\gtrsim\SI{1.2}{MHz}$. This is too fast to be detected given the limited time resolution of \Li~\bnmr. Moreover, the decoherence or any broadening of such oscillations would rapidly depolarize the \Li. Both of these effects would manifest as an immediate loss of polarization at early times (cf.~Ref.~[\onlinecite{Cortie2016}]). However, we do not observe such an instant depolarization of the \Lip\ spin. Note that this argument can exclude both a ferromagnetic state and the presence of a \Li\ induced electronic spin polarization, similar to that observed with muons \cite{Krieger2023PRM}.
An antiferromagnetic state with a vanishing or compensated net local field at the \Li\ site may also be considered. However, at the phase transition one would expect critical fluctuations to be present, which should cross the \Li\ Larmor frequency and lead to a SLR peak at \TN\ (cf.\ 1T-CrSe$_{2}$~\cite{Ticknor2020RSCA}), which would not be suppressed by the low applied field in our measurements. While the SLR peak around~$\sim\SI{15}{K}$ could be caused by magnetic fluctuations, it occurs at a temperature too low to be compatible with the magnetic-like transition observed with \mSR. Together with the absence of any magnetic signal in \Tenmr-NMR\cite{Krieger2023PRM}, this serves as a clear evidence that \tHMoTet\ does not exhibit an intrinsic magnetic order.

\section{Conclusion}\label{sec:Concl} 
We have studied the behavior of isolated \Li\ impurities implanted in \tHMoTet. Using DFT we have identified the \Li\ stopping site at the 2a Wyckoff position. In contrast to implanted muons, we find no theoretical or experimental indications of an intrinsic or \Li\ induced magnetic moment or any type of magnetic order. Instead, we detect \Li\ dynamics above $\sim\SI{200}{K}$, potentially caused by stochastic motion in a local cage. Finally, we have demonstrated that a frequency-comb measurement can reliably extract the quadrupolar splitting, even in the presence of a complex, multi-component line shape.

\section*{Acknowledgements}
The \bnmr\ experiments were performed at the TRIUMF Centre for Materials  and 
Molecular Science. 
We thank G.~D.~Morris, C.~D.~P.~Levy, R.~Li, B.~Hitti, S.~Daviel and R.~Abasalti 
for helpful discussions and technical support. 
J.A.K.~and S.H.~acknowledge
support by the Swiss National Science Foundation (JAK: SNF-Grants
No.~200021\_165910 and No.~P500PT\_203159; SH: No.~200021-159736).
E.V.C.~acknowledges funding by Saint Petersburg State University project
for scientific investigations (ID No. 125022702939-2).
The work at  Warwick 
was supported by EPSRC, UK, through Grant EP/M028771/1~. 
D.F., V.L.K., and J.T.~acknowledge support from a QuEST fellowship. A.C.~and R.M.L.M.~acknowledge 
the support of IsoSiM scholarships. 
Finally, we note that a preliminary version of some of these results was published as part of a PhD thesis, Ref.~[\onlinecite{KriegerThesis}].

\bibliography{Biblio}

\end{document}